\documentstyle[12pt,epsf,epsfig]{article}

\newcommand{\s}{\smallskip}
\def\ltap{\ \raisebox{-.4ex}{\rlap{$\sim$}} \raisebox{.4ex}{$<$}\ }
\def\gtap{\ \raisebox{-.4ex}{\rlap{$\sim$}} \raisebox{.4ex}{$>$}\ }

\begin{document}

\begin{titlepage}
\noindent
\phantom{a}     \hfill         ZU--TH 5/98           \\
\phantom{a}     \hfill         PM--98/1              \\
\phantom{a}     \hfill         GDR--S--011           \\
\phantom{a}     \hfill         hep-ph/9806301        \\[3ex]
\begin{center}

{\bf LOWER BOUNDS ON CHARGED HIGGS BOSONS} \\ 
{\bf FROM LEP AND THE TEVATRON}  \\[5ex] 
{ F. M.\ BORZUMATI$^{\,1,2}$ and A. DJOUADI$^{\,2}$}  \\[1.5ex]            
$^1$ 
{\it Institut f\"ur Theoretische Physik,
     Universit\"at Z\"urich}                       \\
{\it   Winterthurerstrasse 190,
     8057 Zurich, Switzerland}                     \\[1.1ex]
$^2$
{\it Physique Math\'ematique et Th\'eorique,
  Universit\'e de Montpellier}                        \\ 
{\it  F-34095 Montpellier Cedex 5, France}         \\[10ex]
\end{center}
{\begin{center} ABSTRACT \end{center}}
\vspace*{1mm}
{\noindent
We point out that charged Higgs bosons can decay into final states
different from $\tau^+ \nu_\tau$ and $c \bar{s}$, even when they are
light enough to be produced at LEP\,II or at the Tevatron through 
top quark decays. These additional decay modes are overlooked in
ongoing searches even though they alter the existing lower bounds on
the mass of the charged Higgs bosons that are present in
supersymmetric and two Higgs doublets models.}
\vfill
\end{titlepage}

The discovery of a charged Higgs boson would be an unambiguous signal
of an extended Higgs sector and possibly of supersymmetry.  In
supersymmetric models, at least two Higgs doublets are needed to give
mass to all fermions: one is coupled only to down-type quarks and
leptons; the other, only to up-type quarks. A Two Higgs Doublet Model
(2HDM) is said of Type II if the doublets are coupled as in
supersymmetric models with minimal particle content. It is said of
Type~I if one Higgs doublet does not couple to fermions at all and the
other couples as the Standard Model (SM) doublet. \s

After electroweak symmetry breaking, five physical states remain: two
CP-even Higgs bosons $h$ and $H$ (with $m_h < m_H$), a CP-odd Higgs
boson $A$, and two charged states $H^\pm$.  The
charged-Higgses--fermions interactions, can then be comprehensively
expressed as:
\begin{equation} 
{\cal L} =  \frac{\,g}{\sqrt{2}}  \left\{
\left(\!\frac{{m_d}_i}{M_W}\!\right)
      \mbox{\rm{X}} \,{\overline{u}_L}_j V_{ji}  \, {d_R}_i+
\left(\!\frac{{m_u}_i}{M_W}\!\right)
      \mbox{\rm{Y}} \,{\overline{u}_R}_i V_{ij}  \, {d_L}_j 
\, + \left(\!\frac{{m_l}_i}{M_W}\!\right)
      \mbox{\rm{Z}} \,{\overline{\nu}_L}_i \, {e_R}_i 
      \right\} H^+  +{\rm h.c.}\,, 
\label{higgslag}
\end{equation}
where $V$ is the CKM matrix. The equality 
$\mbox{\rm{X}}=\mbox{\rm{Z}}=1/\mbox{\rm{Y}}={\rm tan}\,\beta$, with 
${\rm tan}\,\beta$ the ratio of the two vacuum expectation values,
identifies 2HDMs of Type~II and supersymmetric models;
$\mbox{\rm{Y}}=-\mbox{\rm{X}}=-\mbox{\rm{Z}}= \cot\!\beta$, identifies
2HDMs of Type~I. \s

Besides the mass of $h$, $H$, $A$, and $H^\pm$, two additional
parameters are needed to describe the Higgs sector in 2HDMs of Types~I
and~II: ${\rm tan}\,\beta$ and the mixing angle $\alpha$.  In
supersymmetric models, the Higgs sector is more constrained, and only
two free parameters are needed at the tree level, $m_A$ and ${\rm
tan}\,\beta$.  Supersymmetry induces a relation between ${\rm
tan}\,2\beta$ and ${\rm tan}\, 2 \alpha$ and the well-known tree-level
sum rule $m_{H^\pm}^2=m_W^2+m_A^2$, which is only mildly altered by
one-loop corrections~\cite{HMCORR}.  Together with the experimental
lower bound on $m_h$, $m_A >92\,$GeV, for 
${\rm tan}\,\beta>1$~\cite{JAMBOREE}, this sum rule makes the
supersymmetric charged Higgs bosons possible candidates for discovery
at the Tevatron, but not at LEP~II. \s

Strong constraints on charged Higgs bosons come from searches of
processes where $H^\pm$ is exchanged as a virtual particle. Among
them, the measurement of the inclusive decay 
$\bar{B} \to X_s \gamma$~\cite{CLEO} excludes charged Higgs bosons in
a 2HDM of Type~II up to $\sim 165\,$GeV~\cite{FBCG}; however it is, in
general, inconclusive for supersymmetric models~\cite{BBMR} and 2HDMs
of Type I~\cite{CDGG,FBCG}.  Other indirect bounds on the ratio
$m_{H^\pm}/{\rm tan}\,\beta$ come from inclusive semileptonic
$b$-quark decays $B \to D \tau \nu_\tau$, 
$m_{H^\pm}\gtap 2.2\,{\rm tan}\,\beta\,$GeV~\cite{BCTAUconst} and from
$\tau$-lepton decays, $m_{H^\pm} \gtap 1.5\,{\rm
tan}\,\beta\,$GeV~\cite{TAUconst}.  They apply to charged Higgs bosons
of Type~II in 2HDMs and supersymmetric models. In the former, however,
they are non-competitive with the stronger lower bound due to the
measurement of $\bar{B}\to X_s \gamma$; in the latter they are already
saturated by the above sum rule and the lower bound on $m_A$.
Constraints on the low-${\rm tan}\, \beta$ region and light $H^\pm$ in
Type~I models come from the measurement of $Z\to b \bar{b}$ and
$B^0$--$\bar{B}^0$ mixing (see discussion in~\cite{CDGG}). \s

It is possible that the 2HDMs described above are only ``effective''
models, i.e. the low-energy remnant of Multi-Higgs-Doublets models,
with the same number of degrees of physical states non-decoupled at
the electroweak scale.  In this case, more freedom remains in the
possible values that $\mbox{\rm{X}}$, $\mbox{\rm{Y}}$, and
$\mbox{\rm{Z}}$ can acquire.  For 
$\mbox{\rm{X}} = -1/\mbox{\rm{Y}} =-a$, with $a\ge 2$, for example, 
charged Higgs bosons with $m_{H^\pm} = 100\,$GeV can escape the 
$\bar{B}\to X_s \gamma$ constraint~\cite{FBCG}, while having widths 
for decays into light fermions substantially coinciding with those
obtained in a 2HDM of Type~II.  Moreover, lepton and quark couplings
in~(\ref{higgslag}) may be unrelated, thus rendering the indirect
bounds from $b$-quark and $\tau$-lepton decays independent of that 
coming from $\bar{B} \to X_s \gamma $.  Indirect and direct bounds
are, therefore, all equally necessary in providing the complementarity
that allows the exclusion of certain ranges of $m_{H^\pm}$ in
supersymmetric models, in Type~I and Type~II 2HDMs, and in those
models that may counterfeit them in one specific search. \s

Charged Higgs bosons are searched for at LEP\,II, above the LEP\,I limit, in the
range $45 \ltap m_{H^\pm} \ltap 100\,$GeV and at the Tevatron in the range
$m_{H^\pm} < m_t -m_b$, i.e. when produced by a decaying $t$-quark. Searches at
LEP\,II rely on the assumption that no $H^+$ decay mode, other than $c \bar{s}$
and $\tau^+ \nu_\tau$, is kinematically significant; they give a limit
$m_{H^\pm} \gtap 78.6\,$GeV~\cite{H+LEP}, which applies to 2HDMs of Types~II
and~I. Indeed, within the assumption $BR(H^+ \to c \bar{s},\tau^+ \nu_\tau)
\simeq 100\%$, in Type~I models the two branching ratios are ${\rm
tan}\,\beta$-independent and approximately equal to those obtained in Type~II
models with ${\rm tan}\,\beta=1$. \s

At the Tevatron, searches of an excess of $t\bar t$ events in the
$\tau$ channel provide a ${\rm tan}\,\beta$--$m_{H^\pm}$ exclusion
contour that constrains the very-large-${\rm tan}\,\beta$ region in
supersymmetric models and 2HDMs of Type~II~\cite{CONWAY}, for which
the rate of $t \to H^+ b$ is large.  Similarly large is this rate in
the region of low ${\rm tan}\,\beta$ (${\rm tan}\, \beta \ltap 1$),
for Type~II Yukawa couplings. Searches of $H^+$ apply in this region
to the non-supersymmetric case. They are carried out, specifically for
this type of couplings, looking for: {\it i)} a deficit in the $e$,
$\mu$ channels, due to $H^+\to c \bar s$, for 
$m_{H^\pm} \ltap 130\,$GeV, {\it ii)} a larger number of taggable
$b$-quarks due to $H^+ \to t^\ast b \to \bar{b} b W$ for 
$m_{H^\pm} \gtap 130\,$GeV~\cite{MA,MORIOND}.  Given the limited
luminosity at present available at the Tevatron ($\sim 1$ fb$^{-1}$),
there is no sensitivity to the intermediate range of 
${\rm tan}\,\beta$ where the rate $t \to H^+ b$ becomes low. This
region, partially accessible at the upgraded Tevatron, will be fully
covered at the LHC~\cite{GUNION}.\s

The aim of this letter is to show that there exist additional decay
modes, which are overlooked in ongoing searches of $H^\pm$ within 2HDMs
and supersymmetric models, and which alter the existing lower bounds
on $m_{H^\pm}$.  In the following, the considered type of weak scale
supersymmetry has minimal particle content and R-parity
conservation. No specific assumption is made on the superpartner
spectrum and on the scale/type of messengers for supersymmetry
breaking.  All branching ratios presented for supersymmetric models
are calculated using HDECAY~\cite{HDECAY}. \s

In 2HDMs, these modes are $ H^+ \to A W^+$ and/or $ h W^+$
($HW^+$). They produce mainly the same final state $\bar{b} b W^+$,
as the above-mentioned $\bar{b} t^\ast$ mode and, to a lesser extent,
the state $\tau^+ \tau^- W^+$.
Our statement is based on the fact that there is no stringent lower bound on
$m_A$ and/or $m_h$ coming from LEP~\cite{MK}. Indeed, since the mixing
angle $\alpha$ is, in this case, a free parameter, one can
think of a scenario in which the coupling $ZhA$ vanishes. This coupling
being proportional to $\cos(\beta-\alpha)$, the required direction is
$\alpha=\beta \pm \pi/2$.  In this case, the process $Z^\ast \to hA$
does not occur and the LEP~II bound $m_A>92$ GeV obtained for
supersymmetric models does not hold. Nevertheless, the cross section
for the process $e^+ e^- \to Z^* \to hZ$, proportional to
$\sin^2(\beta-\alpha)$, is not suppressed with respect to that for the
corresponding production mechanism of the SM Higgs boson, and the
LEP~II bound $m_h >114\,$GeV~\cite{JAMBOREE} applies to our case. 
The coupling $ZHA$, still proportional to
$\sin(\beta-\alpha)$, has also full strength,
whereas $H Z Z $ vanishes. The process 
$Z^\ast \to H A$ could in principle provide a bound on $m_A$ depending 
on $m_H$ and ${\rm tan}\,\beta$. For large $m_H$, however, no real lower 
bound can be imposed on $m_A$.  Conversely, even without making specific
choices on the angle $\alpha$, one can assume $h$ to be heavy enough
to render impossible any significant lower bound on $m_A$.
The other two production mechanisms possible at LEP~I (they require
larger numbers of events than LEP~II can provide) are the decay 
$Z\to A\gamma$ and the radiation out of $b\bar b$ and $\tau^+\tau^-$
pairs~\cite{ASTRAHLUNG}. The first is mediated only by fermion loops,
unlike the decay $Z \to h \gamma$, which has additional contributions
from $W$-boson loops. The corresponding rate is about two orders of
magnitude smaller than that for $Z \to h \gamma$ and therefore too
small to allow for a visible signal~\cite{A0}. The second process
allows for sizeable rates only for very large values of
${\rm tan}\,\beta$. No bound can be obtained for non-extreme values of
${\rm tan}\,\beta$ and for 2HDMs of Type~I.  In general, therefore, one
remains with the rather modest bound from the decay 
$\Upsilon\to A \gamma$, which has been searched for by the 
Crystal Ball Collaboration~\cite{CRISTALBALL}, $m_A > 5\,$GeV. \s

\begin{figure}[htbp]
\vspace*{-4.9cm}
\hspace*{-4.3cm}
\mbox{\psfig{figure=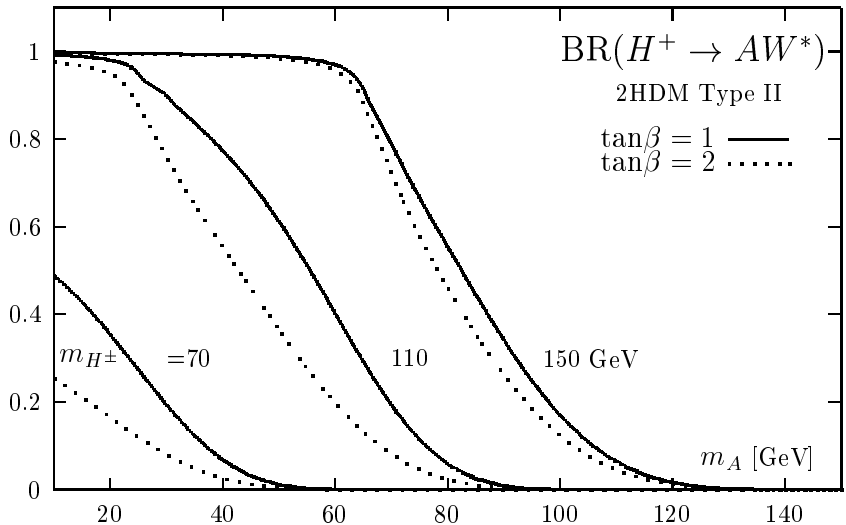,width=20cm}}
\vspace*{-16.5cm}
\caption[f5]{Branching fractions for the decay $H^\pm \to AW^*$ as a 
 function of $m_A$ for three values of $m_{H^\pm}=70,110$ and 150 GeV
 and two values ${\rm tan}\,\beta=1$ (solid) and 2 (dotted).}
\label{wamode}
\end{figure}

If one recalls that the interaction term $H^+ W^-A$ is weighted by a gauge
coupling, unsuppressed by any projection factor, it is clear that the decay
$H^+ \to AW^+$ can be rather important for Type~I models, or for models of
Type~II with small ${\rm tan}\,\beta$. This remains true even for an off-shell
$W$-boson, in spite of the additional propagator and weak coupling that are
then required. For a 2HDM of Type~I and Type~II with ${\rm tan}\,\beta=1$, the
branching ratios $BR(H^+ \to AW^+)$ are shown in Fig.~1 as functions of $m_A$
for different values of $m_{H^\pm}$ (solid lines).  Already for $m_{H^\pm}=
70\,$GeV, roughly the lower bound obtained at LEP~II when $BR(H^+\to c \bar{s},
\tau^+ \nu_\tau) \simeq 100\%$ is assumed, the branching ratio is 50\%--20\%
for $m_A=10$--$30\,$GeV. More strikingly, for heavier $H^\pm$, when the
$W$-boson is not too far from being on shell, this decay mode becomes the
dominant one. We also show in Fig.~1 the branching ratios for this decay mode
in a Type~II model with a higher value $\tan\beta=2$ (for Type~I model, the
situation does not change). $BR(H^\pm \to W^*A)$ is of course smaller because
the competing decay mode, $H^- \to \tau^- \nu_\tau$, has an enhanced decay
width.  This is more striking for low $m_{H^\pm}$ values when the $H^\pm \to
AW^*$ decay channel occurs only at the three--body level. For a heavier $H^\pm$
boson, values of $\tan\beta$ slightly larger than unity do not change the main
trend.  This is particularly true when the $W$ boson is on--shell as in the
example with $M_H \sim 150$ GeV and a light pseudoscalar $A$ boson. In this
case, only for much larger $\tan \beta$ values that the $H^- \to \tau \nu$ 
decay mode becomes dominant and then, the search for the $H^\pm$ boson at 
LEP~II will be the standard one and the limit $m_{H^\pm} \geq 78.6$ GeV 
form $\tau \nu$ and $cs$ decay \cite{H+LEP} will hold (for intermediate 
$\tan\beta$ values, one has to take into account simultaneously all  
decay modes, rendering the analysis more complicated). \s 

Since the two modes $hW^+$ and $HW^+$ are forbidden respectively by our choice
of $\alpha$ and the requirement of a very heavy $H$, the other competing
channels are $\tau^+\nu_\tau$, $c\bar{s}$ for $m_{H^\pm}$ in the LEP~II range,
and $\tau^+\nu_\tau$, $c\bar{s}$, and $\bar{b} t^\ast$ in the Tevatron
searches.  In Fig.~2, the final branching ratio $BR(H^+\to \bar{b} b W^+)$ is
shown as a function of $m_{H^\pm}$ in a 2HDM of Type~II, with our choice of
$\alpha$, for different values of ${\rm tan}\,\beta$ and of $m_A$. For the
larger $m_A$, the mode $AW^+$ is forbidden.  Indeed, above $m_{H^\pm} =
130\,$GeV the mode $c\bar{s}$ is quickly taken over by $\bar{b} t^\ast$, with
the same ${\rm tan}\,\beta $ dependence, but much larger Yukawa couplings,
which can compensate the virtuality of the $t$-quark. The deviations from this
pattern become striking when the mode $AW^+$ starts being allowed. \s

\begin{figure}[htbp]
\vspace*{-4.99cm}
\hspace*{-4.3cm}
\mbox{\psfig{figure=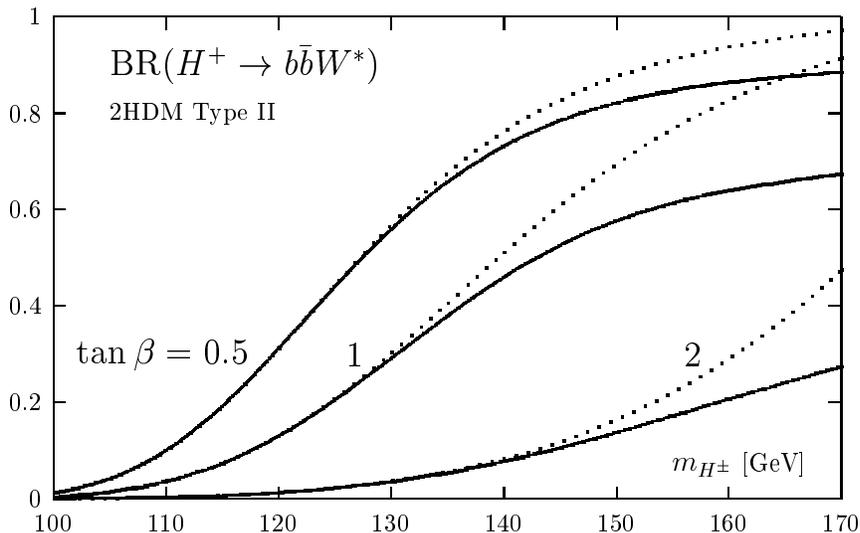,width=20cm}}
\vspace*{-16.6cm}
\caption{Branching fractions for the 
 decay $H^\pm \to \bar{b} b W^+$ as a function 
 of $m_{H^\pm}$ for $m_A=100$ (solid lines) and $200\,$GeV
 (dotted lines) and three different values of ${\rm tan}\,\beta$.}
\label{bbwmode}
\vspace*{-3mm}
\end{figure}
\smallskip

The situation described here corresponds to a particular direction of parameter
space. One could have similarly allowed decays into $hW^+$ and $HW^+$. For
instance, a search strategy based on tagging three $b$-quarks for each produced
$t$-quark at the Tevatron (one $b$--jet coming form the $t \to bH^+$ decay and
two $b$--jets coming from $H^+ \to W^++h,H,A$ with the Higgs bosons decaying
into $b\bar{b}$ pairs) would then sum over all these decays. The corresponding
theoretical branching ratio, however, becomes a function of $m_A$, $m_h$, $m_H$
and $\alpha$, in addition to $m_{H^\pm}$ and ${\rm tan}\, \beta$.  Searches at
LEP~II and the Tevatron aimed at constraining 2HDMs of Type~II in the low ${\rm
tan}\, \beta$ regime and/or 2HDMs of Type~I will have to be modified
accordingly.  Constraints in the region of very large ${\rm tan}\, \beta$ for
Type~II couplings, when only the mode $\tau^+ \nu_\tau$ survives, remain
unchanged. 

In supersymmetric models, and in particular in the minimal version (MSSM),
since $m_A$ cannot be much smaller than $m_{H^\pm}$ and the angle $\alpha$ 
is not an independent parameter, a non-trivial role is played only by the 
mode $ H^+ \to h W^{+\ast}$. However, the branching ratio is large only for 
small values of the parameter $\tan \beta$, $\tan \beta \ltap 2$, for which 
the $h$ boson is constrained to be rather heavy form LEP data \cite{JAMBOREE}
[in fact, such a low $\tan \beta$ scenario is by now excluded].  
For larger values of $\tan \beta$, the $H^+Wh$ coupling is suppressed [and the 
$H^+ \tau \nu$ coupling is enhanced], making the branching ratio for this 
decay mode rather small, not exceeding $\sim 5 \%$ over the LEP allowed region.
[Note that the situation might be different in extensions of the MSSM, 
such as in the case of additional singlet fields, the NMSSM, where $m_{H^\pm}$ 
and $m_A$ are not as strongly related as in the MSSM and the present LEP 
constraints on $m_h$ and $m_A$ do not hold; in this case $BR(H^+ \to hW^*, 
AW^*)$ might be rather large.] \s

In general, however, decays into the lightest chargino $\chi_1^+$ and
neutralino $\chi_1^0$ as well as decays into sleptons are still allowed by
present experimental data, and they dominate when they occur.  (The importance
of the channel $\chi^+_1 \chi^0_1$ for a constrained minimal supersymmetric
model was already discussed in~\cite{FBNPtop}; for decays of MSSM Higgs bosons
into supersymmetric particles, see also Ref.~\cite{SUSYdecays}.) \s

The latest lower bounds on $\chi_1^+$ from LEP~II, $m_{\chi_1^+} \gtap 103.6$
GeV, rely on the assumption of very heavy sleptons and/or a relatively large
mass splitting with the lightest neutralino~\cite{Fabiola}.  For large values
of the Higgs--higgsino mass parameter $\mu$, the lighter chargino and
neutralino states $\chi_1^+$ and $\chi_1^0$ are respectively wino- and
bino-like, with masses $\sim M_2$ and $\sim M_1$. In this case, even assuming
gaugino mass universality at the very high scale: $M_1=\frac{5}{3} {\rm
tan}^2\theta_W M_2\sim \frac{1}{2}M_2$, the decay channel $H^+ \to \chi_1^+
\chi_1^0$ is possible for $m_{H^\pm} > 165\,$GeV. It gives rise to jets or
leptons and missing energy and to $\tau$'s and missing energy.  The branching
ratio $BR(H^+ \to \chi^+_1 \chi^0_1)$ is shown in Fig.~3 as a function of
$m_{H^+}$, for ${\rm tan}\,\beta=4$, $M_2=150\,$GeV and $\mu=200\,$GeV (solid
line). [Here, and in the example for $\tan \beta=4$ in the next discussion, we
have set the sfermion masses at $\sim 1\,$TeV and the trilinear stop coupling
$A_t$ at $\sqrt{6}$ TeV (the so--called maximal mixing scenario) to evade the
experimental bound \cite{JAMBOREE} on the $h$ bound mass.] For these values of
parameters, ${\chi_1^+}$ and ${\chi_1^0}$ have respectively masses of $107$ and
$60\,$GeV.  \s

\begin{figure}[htbp]
\vspace*{-4.9cm}
\hspace*{-4.3cm}
\mbox{\psfig{figure=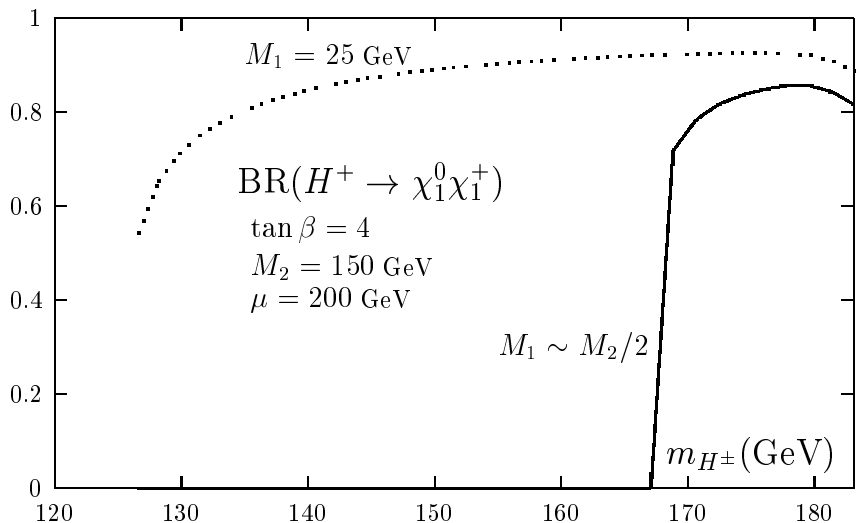,width=20cm}}
\vspace*{-16.8cm}
\caption[]{Branching fractions for the decay $H^+ \to \chi^+_1 \chi^0_1$ 
 as a function of $m_{H^+}$, for ${\rm tan}\,\beta=4$, $M_2=150\,$GeV, 
 $\mu=200$ GeV and two different values of $M_1$: $M_1\sim \frac{1}{2}M_2$ 
 and $M_1=25\,$GeV. All other supersymmetric decay modes are kinematically 
forbidden.}
\vspace*{-3mm}
\end{figure}
\smallskip

The LEP~II limits on $\chi^+_1$ and $\chi^0_1$ become weaker if the assumption
on very heavy slepton masses and/or gaugino mass universality is relaxed. In
both cases, the channel $\chi^+_1 \chi^0_1$ becomes kinematically allowed for
lighter $H^\pm$'s. As an example, we show in Fig.~3 the branching ratio in a
direction of supersymmetric parameter space with $M_1$ disentangled from $M_2$
(dotted line). While keeping all other parameters fixed to the previous values,
$M_1$ is set to $25\,$GeV, which induces a mass for ${\chi_1^0}$ of $\simeq
19\,$GeV. The mode $\chi^+_1 \chi^0_1$ opens now already at $\sim 125\,$GeV. 
Figure~3 clearly shows that, in the region of moderate ${\rm tan}\,\beta$, if
no other decay of $H^+$ into superpartners is possible, the mode $\chi_1^+
\chi_1^0$ can be dominant if it is kinematically allowed. For $m_{H^\pm}
 \simeq 170\,$GeV and ${\rm tan}\,\beta=4$, the contribution of the $\chi_1^+
\chi_1^0$ mode to the  $H^\pm$'s total decay width, indeed, is respectively
78\% and 92\% for $M_1\sim M_2/2$ and $M_1=25$ GeV. An increase of ${\rm
tan}\, \beta$ reduces the branching ratio  $BR(H^+ \to\chi^+_1 \chi^0_1) $,
while a smaller value of $\tan \beta$, if allowed, would make this decay mode
even more dominant, in particular in the case of non--unified gaugino masses.
\s 

The existing lower bounds on the charged slepton masses from LEP~II, are
respectively 95, 88, and $76\,$GeV for $\tilde{e}$, $\tilde{\mu}$,
$\tilde{\tau}$ when the mass difference with the lightest neutralino is rather
large $(\Delta M \gtap 15$ GeV) and the sleptons are assumed to decay
exclusively into $\ell^\pm \chi_1^0$ final states \cite{ALEPH}. These
bounds, in particular in the case of $\tilde{\tau}$, can be much weaker
if they are nearly degenerate with the LSP neutralino. For sneutrinos, an
absolute bound $\gtap\,45\,$GeV comes from the measurement of the invisible $Z$
boson decay width.  Hence, the decay $H^+ \to \tilde{\tau}^+ \tilde{\nu}_\tau$
is therefore kinematically allowed and produces a final $\tau^+$ + missing
energy, but with a softer $\tau^+$ than that coming from the direct decay $H^+
\to {\tau}^+ {\nu}_\tau$. We show in Fig.~4 the relative branching ratio for 
two choices of input parameters: \s

{\it a)} ${\rm tan}\,\beta=4$, $M_2 \sim 2M_1=120$ GeV, $\mu=-500$ GeV, 
$m_{\tilde{l}_L}=m_{\tilde{l}_R}=m_{\tilde{l}}=90\,$GeV and $A_\tau=0$ 
(small or moderate mixing scenario). This 
leads to a slepton spectrum: $m_{\tilde{\nu}} \sim 66$ GeV, $m_{\tilde{e}} 
\sim m_{\tilde{\mu}} \sim$ $100 \,$GeV and the two $\tilde{\tau}$ masses
$\sim 20$ GeV below and above this value (the lightest chargino and neutralino 
masses are $m_{\chi_1^+} \sim 123$ GeV and $m_{\chi_1^0} \sim 60$ GeV). \s

{\it b)} ${\rm tan}\,\beta=25$, $M_2 \sim 2M_1 =\mu=150$ GeV, $m_{\tilde{l}}=
100\,$GeV and $A_\tau=-800\,$GeV (the large $A_\tau$ value is chosen to 
maximize the $H^\pm \tilde{\tau}\tilde{\nu}_\tau$ coupling as will discussed 
later). This leads to the following spectrum: 
$m_{\tilde{\nu}} \sim 76$ GeV, $m_{\tilde{e}} 
\sim m_{\tilde{\mu}} \sim$ $110 \,$GeV and the $\tilde{\tau}_1$ mass
$m_{\tilde{\tau}_1} \sim 63\,$GeV almost degenerate with the lightest
neutralino mass $m_{\chi_1^0} \sim 61$ GeV (therefore the decay 
$\tilde{\tau}_1 \to \chi_1^0 \tau$ gives very soft $\tau$ leptons, which
will be overwhelmed by the $\gamma \gamma$ background and the LEP\,II lower 
limit on $m_{\tilde{\tau}_1}$ does not hold in this case). \s

Below the threshold for scenario a) with $\tan \beta=4$, the dominant decays 
are $\tau^+ \nu_\tau$ and $hW^*$, while $AW^+$ and $c\bar{s}$ are below the 
percent level. Above the threshold, the branching ratio for the decay
$H^\pm \to \tilde{\tau}^\pm \tilde{\nu}_\tau$ can become rather sizeable,
possibly reaching the level of $\sim 30\%$.  For large enough $H^\pm$ masses,
the channels $H^\pm \to \tilde{\mu}^\pm \tilde{\nu}_\mu$ and $\tilde{e}^\pm 
\tilde{\nu}_e$, open up, leading to an increase of $BR(H^\pm \to \tilde{\ell}
\tilde{\nu})$ up to $\sim 80\%$. In scenario b) with $\tan \beta=25$ and a
large $A_\tau$ value, $\tau^+ \nu_\tau$ decays are by far dominant below the 
threshold. When the decay $H^\pm \to \tilde{\tau}^\pm \tilde{\nu}_\tau$ opens 
up, the branching ratio quickly reaches the level of $\sim 75\%$. \s

The prominence of $\tilde{\tau}^+ \tilde{\nu}_\tau$ decays observed above
threshold is explained by the $H^\pm$ coupling to sleptons. For small stau
mixing and small $\tan \beta$ values, the Lagrangian term $H^+\tilde{\nu}_L^*
\tilde{l}_L$, $- (g/\sqrt{2}) M_W \sin 2 \beta$, is very large with respect to
the Yukawa coupling $-(g/\sqrt{2}) (m_\tau/M_W) \tan\beta$.  Owing to the
$\sin2 \beta$ dependence, this term quickly dies off for increasing ${\rm
tan}\, \beta$. In this case, however, there exists other directions of
parameter space where this decay mode still has a branching ratio $\sim 100\%$.
For instance, when $A_\tau$ and $\tan \beta$ are large,  since the coupling 
of the Lagrangian term $H^+\tilde{\nu}_L^* \tilde{\tau}_R$: $-(g/\sqrt{2}) 
(m_\tau/M_W) (\mu+ A_\tau {\rm tan}\,\beta)$ becomes very strong, the decay 
rate is enhanced as shown in Fig.~4 (note that for $A_\tau \sim \mu \,{\rm 
tan}\,\beta$, the left--right mixing in the slepton mass matrix tends to 
vanish). \s

\begin{figure}[htbp]
\vspace*{-4.9cm}
\hspace*{-4.3cm}
\mbox{\psfig{figure=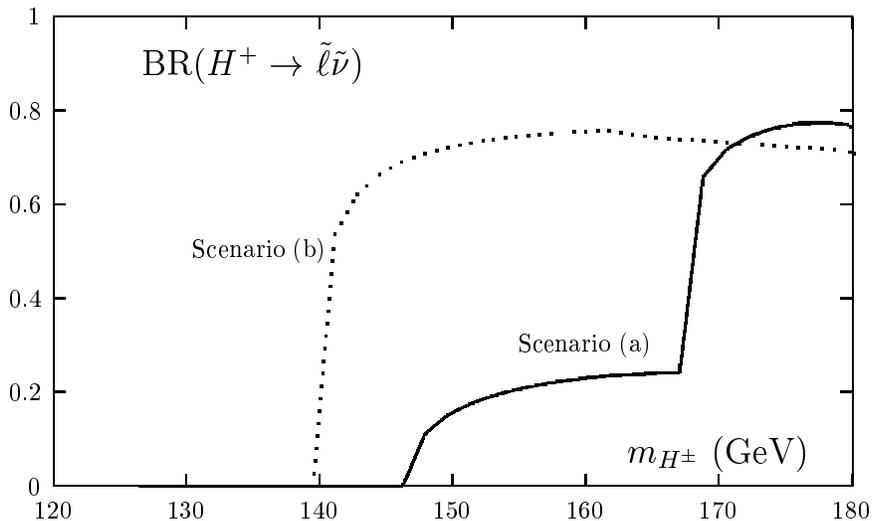,width=20cm}}
\vspace*{-16.8cm}
\caption[]{Branching fractions for the decay 
 $H^+ \to \tilde{\ell}^+ \tilde{\nu_\ell}$ as a function of $m_{H^+}$, 
 for the two different sets of supersymmetric parameters a) and b) given 
in the text.} \label{stsnmode}
\vspace*{-5mm}
\end{figure}
\bigskip

Summarizing, at very large ${\rm tan}\, \beta$, a possible excess of $\tau$'s
softer than those predicted by a 2HDM of Type~II may indicate the presence of a
heavier $H^\pm$ decaying into $\tilde{\tau}^+\tilde{\nu}_\tau$.  Searches in
the region of ${\rm tan}\, \beta \gtap 1$ should already consider multi-$b$
signals coming from $hW^{+\ast}$, $\bar{b}bW^+$ as well as $\tau$-signals with
a wide momentum distribution coming from $\chi^+_1 \chi^0_1$, $\tilde{\tau}^+
\tilde{\nu}_\tau$, and $\tau^+ \nu_\tau$ and jets/leptons + missing energy
signals from $\chi^+_1 \chi^0_1$.  \s

It is needless to say that all these modes will play an important role in 
future searches and will not be blind to the intermediate range of ${\rm 
tan}\,\beta$. This would be
particularly the case at the Tevatron Run\,II where the $H^\pm$ bosons, if
light enough, can be produced copiously in top quark decays (other production
channels would have much smaller rates) \cite{Tevatron}. While it would be
always possible to detect them in a ``disappearance" search (i.e. by looking at
one top quark decaying into the standard mode, $t \to W^+b$, which should have
a relatively large branching ratio, and ignoring the decay products of the
other) \cite{Tevatron}, the direct search for 2HDMs $H^\pm$ bosons decaying
into $Wb\bar{b}$ final states would be in principle relatively easy with high
enough luminosity, since the performances of the CDF and D0 detectors for
$b$--quark tagging are expected to be rather good. In the case of SUSY models,
where the $H^\pm$ should be tagged though the leptonic decays of charginos or
$\tau$ sleptons, the detection might be more challenging because of the
softness of these particles. A detailed Monte--Carlo analysis, which is beyond
the scope of this letter, will be needed to assess the potential of the
Tevatron to search for the $H^\pm$ bosons in these new decay channels. 
\bigskip

\noindent {\bf Note Added:} After the first submission of this paper, a search
for 2HDM charged Higgs bosons decaying into $AW^*$ finals states has been
performed by the OPAL collaboration \cite{H+WA}; constraints in the $(m_A, 
m_{H^\pm})$ plane for various $\tan \beta$ values have been set. In addition,
the decay mode $H^\pm \to b\bar{b}W^\pm$ has been taken into account in
simulations of $H^\pm$ searches at the upgraded Tevatron \cite{Tevatron}  
and at the LHC \cite{Ketevi}. Some of the decays modes discussed here 
have been also revisited in theoretical papers in the context of 2HDM 
\cite{pap2HDM} and the MSSM \cite{papMSSM}. \bigskip

\noindent {\bf Acknowledgments:} The authors thank J.~Conway, M.~Drees,
E.~Duchovni, L.~Duflot, M.~Felcini, E.~Gross, P.~M\"attig, and N. Polonsky for
discussions.  F.~B. acknowledges the hospitality of the CERN theory division
during the completion of this paper. This work was supported in part by the
Schweizerischer Nationalfonds and CNRS (France). 



\begin{thebibliography}{99}

\bibitem{HMCORR} H.E.~Haber, R.~Hempfling and A.H.~Hoang, Z. Phys C75 (1997) 
539 and references therein.
%
\bibitem{JAMBOREE} The LEP Higgs working group, hep-ex/0107029 and 
hep-ex/0107030. 
%
\bibitem{CLEO} M.S.~Alam {\it et al.} (CLEO Coll.),  Phys. Rev. Lett. 74 
(1995) 2885.
%
\bibitem{FBCG} F.M.~Borzumati and C.~Greub,  Phys. Rev. D58 (1998) 074004 and  
D59 (1999) 057501; hep-ph/9810240.
%
\bibitem{BBMR} S.~Bertolini, F.~Borzumati, A.~Masiero and G.~Ridolfi,
Nucl. Phys. B353 (1991) 591;  \\
F.M.~Borzumati, Z. Phys C63 (1994) 291.  
%
\bibitem{CDGG} M.\,Ciuchini, G.\,Degrassi, P.\,Gambino and G.F.\,Giudice,
Nucl. Phys. B527 (1998) 21.
%
\bibitem{BCTAUconst} K.~Kiers and A.~Soni, Phys. Rev. D56 (1997) 5786 and 
references listed there.
%
\bibitem{TAUconst} A.~Stahl and H.~Voss, Z. Phys C74 (1997) 73. 
%
\bibitem{H+LEP} LEP Higgs Working Group, hep-ex/0107031. 
%
\bibitem{CONWAY} F.~Abe {\it et al.} (CDF Coll.), Phys. Rev. Lett. 79 (1997) 
357. 
%
\bibitem{MA} A.~Djouadi, J.~Kalinowski and P.~Zerwas, Z. Phys. C70 (1996) 427;
\\ E. Ma, D.P.~Roy and J.~Wudka, Phys. Rev. Lett. 80 (1998) 1162.   

\bibitem{MORIOND}
 R.~Hughes, in ``Rencontres de Moriond: Electroweak 
 interactions and Unified Theories'', Les Arcs, France, March 1998.

\bibitem{GUNION}
J.F.~Gunion, A.~Stange, and S.~Willenbrock, hep-ph/9602238, in 
``Electroweak Symmetry Breaking and New Physics at the TeV scale'', 
T.L.~Barklow (Ed.).

\bibitem{HDECAY} A. Djouadi, J.~Kalinowski, and M.~Spira,
  Comput.~Phys.~Commun.~108~(1998)~56.

\bibitem{MK}
 M.~Krawczyk, hep-ph/9612460; \\
 M.~Krawczyk, J.~Zochowski, and P.~M\"attig, Eur. Phys. J. C8 (1999) 495.

\bibitem{ASTRAHLUNG} 
 A.~Djouadi, P.~Zerwas, and Z.~Zunft, Phys. Lett. B259 (1991) 175;
A. Djouadi, J. Kalinowski and P.M. Zerwas Z.~Phys.~C54 (1992) 255
and Mod.~Phys.~Lett.~A7 (1992) 1765; 
J. Kalinowski and M. Krawczyk, Phys.~Lett.~B361 (1995) 66.

\bibitem{A0} 
 See: ``Z Physics at LEP'', CERN Report 89--08, 
 G.~Altarelli, R.~Kleiss, and C.~Verzegnassi (Eds.).   

\bibitem{CRISTALBALL} 
 D.~Antreasyan {\it et al.} (Crystal Ball Coll.),
  Phys. Lett. B251 (1990) 204. 

\bibitem{FBNPtop}
 F.M.~Borzumati, N.~Polonsky, hep--ph/9602433, in 
  ``$e^+e^-$ Collisions at TeV Energies: The Physics Potential'',
 P.M.~Zerwas (Ed.), DESY 96--123D;  \\
 E.~Accomando {\it et al.}, Phys. Rep. 299 (1998) 1.               


\bibitem{SUSYdecays} H. Baer et al., Phys. Rev. D36 (1987) 1363; 
K. Griest, H.E. Haber, Phys. Rev. D37 (1988) 719; 
J.F. Gunion and H.E. Haber, Nucl. Phys. B307 (1988) 445; 
A.~Djouadi et al., Phys. Lett. B376 (1996) 220,
A. Djouadi, J. Kalinowski and P. Zerwas, Z. Phys. C57 (1993) 569 
and Z. Phys. C 74 (1997) 93;  
A.~Bartl et al., Phys. Lett. B389 (1996) 538;
A. Djouadi, Mod.~Phys.~Lett.~A14 (1999) 359 and  
Phys. Lett. B435 (1998) 101.

 \bibitem{Fabiola} F. Gianotti, Talk given at IECHEP, Budapest Jul 2001. 

\bibitem{ALEPH} ALEPH Collaboration (A. Heister et al.), Phys.\,Lett.\,B526
(2002) 206. 

\bibitem{Tevatron}  M. Carena et al., Report of the Higgs working group for 
``RUN II at the Tevatron",  hep-ph/0010338.

\bibitem{H+WA} OPAL Collaboration, Physics Note PN472, contribution to IECHEP
Budapest, July 2001. 

\bibitem{Ketevi} K. A. Assamagan, Y. Coadou and A. Deandrea, ``ATLAS discovery 
potential for a heavy charged Higgs boson", hep-ph/0203121. 

\bibitem{pap2HDM} L. Brucher and R. Santos, Eur.\,Phys.\,J.\,C12 (2000)87; \\
A.G. Akeroyd, A. Arhrib and E. Naimi, Eur.\,Phys.\,J.\,C12 (2000) 451; \\
A.G. Akeroyd, S. Baek, Gi-Chol Cho and K. Hagiwara, hep-ph/0205094. 
 
\bibitem{papMSSM} M.\,Bisset, M.\,Guchait and S.\,Moretti, Eur.\,Phys.\,J.\,C19
(2001) 143; \\ Filip Moortgat (CMS Collaboration), talk given at the meeting 
``Euro--GDR Supersymmetry", Durham,  April 2002. 


\end{thebibliography}
\end{document}